\documentstyle[12pt,epsf]{article}

\newcommand\as{\alpha_S}

\newcommand\MeV{{\,\sl MeV}}
\newcommand\GeV{{\,\sl GeV}}
\newcommand\slv{v\kern-5pt\raise1pt\hbox{$\scriptstyle/$}\kern1pt}
\newcommand\spur{\mathop{\rm Tr}\nolimits}

\begin{document}

\thispagestyle{empty}
\begin{flushright}
MZ-TH-96-26\\
August 1996\\
\end{flushright}

\thispagestyle{empty}
\vspace{0.5cm}
\begin{center}
{\Large\bf Heavy Baryons and QCD Sum Rules}\\[.5cm]
O.I. YAKOVLEV \footnote{\offinterlineskip
Yakovlev@dipmza.physik.uni-mainz.de\\
Invited talk given at the III German-Russian Workshop on Heavy Quark 
Physics, Dubna, Russia, \\
May 20-22, 1996 to appear in the Proceedings}\\[.5cm]
{\it Budker Institute of Nuclear Physics (BINP),\\[.1cm]
pr. Lavrenteva 11, Novosibirsk, 630090, Russia\\[.1cm]
and\\[.1cm]
Institut f\"ur Physik, Johannes-Gutenberg-Universit\"at,\\[.2cm]
Staudinger Weg 7, D-55099 Mainz, Germany\\}
\vspace{.5cm}
\end{center}

\begin{abstract}\noindent   
We discuss an application of QCD sum rules to the heavy baryons 
$\Lambda_Q$ and $\Sigma_Q$.
The predictions for the masses of heavy baryons, residues and 
Isgur-Wise function are presented.
 The new results on two loop anomalous dimensions 
of baryonic currents and QCD radiative corrections 
(two- and three- loop contributions) to the first two Wilson coefficients 
in OPE are explicitly presented.  
\end{abstract}

\section{Introduction}

Baryons with one $b$ quark are very nice systems for application 
of Heavy Quark Effective Theory (HQET)\cite{HQET}. 
It allows one to organize the determination of the properties of baryons 
in an $1/m_Q$-expansion, the leading term of which gives rise to the 
spin-flavour symmetry. 
Well known predictions of the HQET are the relations between different 
hadron transition form factors. For example, the six form-factors describing 
the $\Lambda_b\to\Lambda_c$ electro-weak transitions are reduced to one 
universal Isgur-Wise function $\xi (v\cdot v')$ 
in the HQ limit~\cite{Georgi,Mannel,Hussein}.
 However, one still remains with many non-perturbative parameters, such as 
the mass of the baryon, the Isgur-Wise function and the averaged kinetic 
and magnetic energy of the baryon, which should be estimated by using some 
non-perturbative method as sum rules~\cite{Vain}, lattice calculation or 
some potential model. 

In the present paper 
we review an application of the QCD sum rule method 
to calculate the masses, residues of heavy baryons and baryonic 
Isgur-Wise function\cite
{Shur,Block,BaDo,GrYa,GrYa1,BaDo1,Cola,Cola1,DaHu,DaHu1,GKY,GKY1,GKY2}. 

\section {\bf Currents, anomalous dimensions and residues}

The currents of the heavy baryon $\Lambda_Q$ and the heavy quark spin 
baryon doublet $\{\Sigma_Q,\Sigma_Q^*\}$ are associated with the 
spin-parity quantum 
numbers $j^P=0^+$ and $j^P=1^+$ for the light diquark system
with antisymmetric and symmetric flavor structure, respectively. 
Adding the heavy quark to the light quark system, one obtains 
$j^P=\frac12^+$ for the $\Lambda_Q$ baryon and the pair of degenerate 
states $j^P=\frac12^+$ and $j^P=\frac32^+$ for the baryons $\Sigma_Q$ 
and $\Sigma_Q^*$. 
The general structure of 
the heavy baryon currents has the form (see e.g.~\cite{GrYa} and refs.\ 
therein)
\begin{equation}\label{current}
J=[q^{iT}C\Gamma\tau q^j]\Gamma'Q^k\epsilon_{ijk}.
\end{equation}
Here the index $T$ means transposition, $C$ is the charge conjugation 
matrix with the properties $C\gamma^T_\mu C^{-1}=-\gamma_\mu$ and 
$C\gamma^T_5C^{-1}=\gamma_5$, $i,j,k$ are colour indices and $\tau$ is a 
matrix in flavour space. The effective static field of the heavy quark is 
denoted by~$Q$. For each of the ground state baryon currents there are two 
independent current components $J_1$ and $J_2$: 
\begin{eqnarray}
J_{\Lambda1}&=&[q^{iT}C\tau\gamma_5q^j]Q^k\varepsilon_{ijk},\qquad
J_{\Lambda2}\ =\ [q^{iT}C\tau\gamma_5\gamma_0q^j]Q^k\varepsilon_{ijk},
  \nonumber\\[7pt]
J_{\Sigma1}&=&[q^{iT}C\tau\vec\gamma q^j]
  \cdot\vec\gamma\gamma_5Q^k\varepsilon_{ijk},\qquad
J_{\Sigma2}\ =\ [q^{iT}C\tau\gamma_0\vec\gamma q^j]
  \cdot\vec\gamma\gamma_5Q^k\epsilon_{ijk}.
\end{eqnarray}

The anomalous dimensions of the heavy baryon currents up to second order was 
calculated recently in~\cite{GKY} and read (for example in the MS scheme 
with naively anticommuting $\gamma_5$)
\begin{eqnarray}
\gamma_{\Lambda1}
  =-8\left(\frac{\alpha_s}{4\pi}\right)+\frac19(16\zeta(2)+40N_F-796)
  \left(\frac{\alpha_s}{4\pi}\right)^2,\label{andimlam1}\\ 
\gamma_{\Lambda2}
  =-4\left(\frac{\alpha_s}{4\pi}\right)+\frac19(16\zeta(2)+20N_F-322)
  \left(\frac{\alpha_s}{4\pi}\right)^2,\label{andimlam2}\\
\gamma_{\Sigma1}
  =-4\left(\frac{\alpha_s}{4\pi}\right)+\frac19(16\zeta(2)+20N_F-290)
  \left(\frac{\alpha_s}{4\pi}\right)^2,\\
\gamma_{\Sigma2}
  =-\frac83\left(\frac{\alpha_s}{4\pi}\right)+\frac1{27}(48\zeta(2)+8N_F+324)
  \left(\frac{\alpha_s}{4\pi}\right)^2. 
\end{eqnarray}
The anomalous dimensions are ingredient of renormalization group invariant 
sum rules. Numerically 
effects of $\gamma_2$ are very small and gives about $1-2\%$ corrections to 
the current redefinition.

Let us define residues $F_i(\mu)$ 
($i=\Lambda,\Sigma$) of the baryonic currents according to   
\begin{eqnarray} \label{residue}
\langle 0|J(\mu)|\Lambda_Q\rangle=F_{\Lambda}(\mu)u,\qquad
\langle 0|J(\mu)|\Sigma_Q\rangle=F_{\Sigma}(\mu)u, 
\end{eqnarray}
where $u$ is  spinor. 
Then we define bound state energy $E_R$ of the baryon 
$R$ and use the formula
\begin{equation}
m_{\rm baryon}=m_Q+E_{R}+\frac{c}{m_Q}+O(1/m^2_Q),
\end{equation}
 where $m_Q$ is the running heavy quark mass 
at the scale of the same mass 
and $c$ is a coefficient connected with the averaged kinetic and magnetic 
energy in the baryonic state. The coefficient $c$ 
was recently estimated in \cite{Cola} by using QCD sum rules.

\section{Correlator of two baryonic currents}
 
In order to obtain information about the value of the mass of the 
baryon and its residue one considers the correlator 
of two baryonic currents
\begin{equation}\label{correlator}
\Pi(\omega=k\cdot v)=i\int d^4xe^{ikx}\langle 0|TJ(x)\bar J(0)|0\rangle,
\end{equation}
where $k_\mu$ and $v_\mu$ are the residual momentum and four velocity in 
$p_\mu=m_Qv_\mu+k_\mu$, respectively.
$P(\omega)$ can be factorized 
into a spinor dependent piece and scalar function $P(\omega)$ according to
\begin{eqnarray}
\Pi(\omega)=\Gamma'\frac{1+\slv}2\bar\Gamma'
  \frac14\spur(\Gamma\bar\Gamma)2\spur(\tau\bar\tau)P(\omega).
\end{eqnarray}

The correlator function $P_{\rm OPE}(\omega)$ 
satisfies the dispersion relation
\begin{equation}\label{dispersion}
P_{\rm OPE}(\omega)=\int_0^\infty
  \frac{\rho(\omega')d\omega'}{\omega'-\omega-i0}
  +\hbox{\rm subtraction},
\end{equation}
where $\rho(\omega)={\sl Im}(P(\omega))/\pi$ is the spectral density. 

 Following the standard QCD sum rules method~\cite{Vain} the correlator is 
calculated in the region $-\omega\approx 1-2\GeV$, including perturbative 
and non-perturbative effects, where non-perturbative effects can be quite 
important. The non-perturbative effects are taken into account with the help 
of an operator product expansion.
The OPE of two point correlator was discussed in Refs.
\cite{Shur,Block,BaDo,GrYa,BaDo1}. $1/m_Q$ corrections to two and 
three point correlators were calculated in
\cite{DaHu,DaHu1,Cola}. 
The leading perterbative term 
and the next-to-leading term in OPE (gluon condensate contribution)
give the spectral densities
\begin{eqnarray}
\rho_0(\omega)=\frac{\omega^5}{20\pi^4},\quad
\rho_4(\omega)=c\frac{\alpha_s\langle GG\rangle}{32\pi^3}\omega.
\end{eqnarray}
Next we consider the radiative corrections to the 
spectral density of the perturbative contribution. 
There are four different graphs 
contributing to the correlator, which are shown in Fig.~1. 
The fact that all graphs in Fig.~1 have 
two-point two-loop subgraphs greatly simplifies the calculational task. 
One can first evaluate the respective subgraphs which leaves one with a 
one-loop integration. The first two-loop integration can be performed by 
using algebraic method described in~\cite{BrGr}. 
It is important to note 
that the results of the two-loop integration are polynomials of the 
external momentum relative to this subgraph. Hence, the last integration 
is really a one-loop one, but the power of one of the propagators becomes 
a non-integer number.
The evaluation of results in a 
Taylor expansion in $1/\epsilon$ gives \cite{GKY2}
\begin{eqnarray}\label{results1}
P(\omega)\!\!&=&\!\!-\frac{32\omega^5}{(4\pi)^4}\Bigg[
  \left(\frac{-2\omega}{\mu}\right)^{-4\epsilon}
  \frac1{40}\Bigg(\frac1{\epsilon}+\frac{107}{15}\Bigg)\\ \nonumber&&
  +\frac{\as}{4\pi}\left(\frac{-2\omega}{\mu}\right)^{-6\epsilon}
  \Bigg(\frac{n^2-4n+6}{45\epsilon^2}
  +\frac{40\zeta(2)+61n^2-234n+396}{225\epsilon}
  +\frac{(n-2)s}{90}\\ \nonumber&&
  +\frac{5(195n^2-780n+1946)\zeta(2)-2200\zeta(3)
  +4907n^2-18408n+34352}{2250}\Bigg)\Bigg].
\end{eqnarray}
Here 
we assume that matrix $\Gamma_1$ in Eq.(1) is an antisymmetrized 
product 
of $n$ Dirac matrices: $\Gamma=\gamma^{[\mu_1}\cdots\gamma^{\mu_n]}$. 
This introduces a 
$n$- and $s$-dependence in the QCD corrections due to the identities
\begin{equation}\label{Dirac}
\gamma_\alpha\Gamma\gamma_\alpha=h\Gamma=(-1)^n(D-2n)\Gamma,\qquad
  \gamma_0\Gamma\gamma_0=(-1)^ns\Gamma.
\end{equation}       
The correlator is renormalized by the renormalization factor of the
baryonic current, which is derived in~\cite{GrYa},
\begin{equation}\label{oneloopAD}
P(\omega)=Z^2_JP^{\rm ren}(\omega)\qquad\hbox{\rm with\ }
Z_J=1+\frac{\as C_B}{4\pi\epsilon}(n^2-4n+6).
\end{equation}
The multiplication with $Z^2_J$ results in the cancelation of the second 
power in $1/\epsilon$. The first power in $1/\epsilon$ is purely real and 
hence does not contribute to the spectral density. The renormalized 
spectral density $\rho^{\rm ren}(\omega)={\sl Im}(P^{\rm ren}(\omega))/\pi$ 
must in fact be finite and can be read immediately off Eq.~(\ref{results1}),
\begin{eqnarray}\label{ro}
\rho^{\rm ren}(\omega)&=&\rho_0(\omega)\Bigg[
  1+\frac{\as}{\pi}r(\omega/\mu)\Bigg],\quad\mbox{where}\quad
  \rho_0(\omega)\ =\ \frac{\omega^5}{20\pi^4}\quad\mbox{and}\\
r(\omega/\mu)&=&\frac23(n^2-4n+6)\ln\left(\frac\mu{2\omega}\right)
  +\frac2{45}(60\zeta(2)+38n^2-137n+273).\nonumber
\end{eqnarray}
The $\as$ correction can be seen to depend on the properties of the 
light-side Dirac matrix in the heavy baryon current. 
The results in the naively anticommuting $\gamma_5$-scheme (AC) are
\begin{eqnarray}\label{ro1}
r_{\Lambda 1}(\omega /\mu)\!\!&=&
\!\!4\ln\left(\frac\mu{2\omega}\right)
+\underbrace{\frac{2(20\zeta(2)+91)}{15}}_{\approx 16.51},\\ \nonumber
r_{\Lambda 2,\Sigma 1}(\omega /\mu)
&=&2\ln\left(\frac\mu{2\omega}\right)
+\underbrace{\frac{4(10\zeta(2)+29)}{15}}_{\approx 12.12},\\ \nonumber
r_{\Sigma 2}(\omega /\mu)\!\!
&=&\!\!\frac23ln\left(\frac\mu{2\omega}\right)
+\underbrace{\frac{2(60\zeta(2)+151)}{45}}_{\approx 11.10}. 
\end{eqnarray}
The coefficient of the logarithmic term $\ln(2\omega/\mu)$ in Eq.~(\ref{ro}) 
coincides with twice the one-loop anomalous dimension given in 
Eq.~(\ref{oneloopAD}), as expected. The results for the Baryons $\Lambda_1$ 
and $\Lambda_2$ in the 't~Hooft-Veltman $\gamma_5$-scheme (HV) differ from 
those presented above. But we know that currents in different schemes 
should be connected by finite renormalization factors $Z$,
$
J_{AC}=ZJ_{HV}.
$
Such factors was recently derived from two-loop anomalous dimensions of 
baryonic currents \cite{GKY}. They read
\begin{equation}Z_{\Lambda 1}=1-\frac{4\as}{3\pi}\quad\mbox{and}\quad
 Z_{\Lambda 2}=1-\frac{2\alpha_S}{3\pi}.
\end{equation} 
By multiplying the results in the 't~Hooft-Veltman scheme by $Z^2$ we 
obtain the same results as in the naively anticommuting $\gamma_5$-scheme.

The results show that the $\as$-corrections amount to about 100\%, which 
makes perturbative QCD radiative ``corrections'' very important in QCD sum 
rules. We used $\mu = 2\omega\approx 1 \GeV$ and 
$\as(\mu)=0.3$. 

\section{Contribution of quark condensate}

Next, we consider the contribution of the quark condensate, which 
appears in the nondiagonal sum rules. The leading and next-to-leading 
spectral density are 
\begin{eqnarray}
\rho(\omega)=-\frac{\langle\bar qq\rangle}{\pi^2}\omega^2
\qquad
\rho_5(\omega)=2\Big(1-\frac c2\Big)
  \frac{\langle\bar qq\rangle m^2_0}{16\pi^2}.
\end{eqnarray}
The radiative corrections to the quark condensate term can be quite 
important, and we take it into account. There are 8 different graphs 
contributing to the correlator, which are shown in Fig.~2. We used the 
same method as we used for the perturbative term. 
The correlator must be renormalized by the renormalization factor of the
baryonic current~\cite{GrYa} and quark condensate,
\begin{eqnarray}
P(\omega)=Z_{J_1}Z_{J_2}Z_{\bar q q}P^{\rm ren}(\omega)\qquad\hbox{\rm with\ }
Z_{J_1}=1+\frac{\as C_B}{4\pi\epsilon}(n^2-4n+6), \\
Z_{J_2}=1+\frac{\as C_B}{4\pi\epsilon}((n-2+s)^2+2),\qquad
  Z_{\bar q q}=1+\frac{\as}{4\pi\epsilon}(-3).  
\end{eqnarray}
This procedure cancels the second power in $1/\epsilon$ while the first power 
in $1/\epsilon$ is purely real and does not contribute to the spectral density.
For the renormalized spectral density we get
\begin{eqnarray}
\rho^{\rm ren}(\omega)
  &=&-\frac{\langle\bar qq\rangle^{\rm ren}\omega^2}{\pi^2}
  \Big[1+\frac{\alpha_s}{4\pi}\Big(\frac43(2n^2-8n+7+2(n-2)s)
  \ln\left(\frac\mu{2\omega}\right)\nonumber\\&&
  +\frac23(8n^2-28n+37+8ns-14s+8\zeta(2))\Big)\Big].
\end{eqnarray}
 Then we proceed with the usual QCD sum rules analysis and equate the 
theoretical result for $P_{\rm OPE}(\omega)$ 
with the dispersion integral over the hadron states saturated by the 
lowest lying state with the bound state energy $E_R$ plus exited states 
and continuum. 
So the phenomenological part of the sum rule is given 
by the spectral density
$
\rho_{\rm ph}(\omega)=\rho_{\rm res}(\omega)+\rho_{\rm cont}(\omega),
$
where the contribution $\rho_{\rm res}$ of the low-lying baryon state is 
$
\rho_{\rm res}(\omega)=\frac{|F|^2}{2}\delta(\omega-E_R).
$
As is usual for the contribution of excited states and continuum 
contributions we assume hadron-parton duality and take 
$\rho_{\rm cont}(\omega)=\theta(\omega-E_C)\rho(\omega)$, where $\rho$ is 
the result of the OPE calculations. 
We apply the Borel transformation 
\begin{equation}
\hat B_T=\lim\frac{\omega^n}{\Gamma(n)}\left(-\frac{d}{d\omega}\right)^n
  \qquad n,-\omega\to\infty,\quad T=-\omega/n\hbox{\rm\ fixed} 
\end{equation}
and finally obtain the sum rule
\begin{equation}\label{sumrule}
\frac12F^2e^{-E_R/T}=\int_0^{E_C}\rho(\omega')e^{-\omega'/T}d\omega'
  +\hat BP_{\rm pc}(T).
\end{equation}

\section{Numerical results}

Let us discuss the sum rule analysis. 
First, we analyse the dependence of 
the bound state energy $E_R$ as a function 
of the energy of continuum $E_C$ and the Borel parameter $T$ in a wide 
region of these arguments and try to find regions of stability in $T$ and 
$E_c$. Second, we fix the energy of continuum $E_c=E_c^{\rm best}$ by 
requiring that $E_R(T)$ should have the highest possible stability in its 
dependence on the Borel parameter $T$. 
We take into account the $\as$-correction to the spectral density and 
obtain
\begin{equation}
E_{\Lambda}=0.9\pm 0.2\GeV \qquad\hbox{\rm and}\qquad
  E_C^{\rm best}=1.2\pm 0.2\GeV.
\end{equation}
The analysis of sum rules at $E_{\Lambda}=0.9\GeV$ gives for the 
$\Lambda$-residue 
\begin{equation}
F_{\Lambda}=(0.03-0.04)\GeV^3\qquad\hbox{\rm and}\qquad E_C^{}=1.0-1.4\GeV.
\end{equation} 
Doing the same analysis for the $\Sigma$ baryon and taking into account the 
$\as$-corrections we obtain
\begin{equation}
E_{\Sigma}=1.1\pm 0.2\GeV\qquad\hbox{\rm and}\qquad E_C^{\rm best}=1.4\pm0.2\GeV.
\end{equation} 
\begin{equation}
F_{\Sigma}=(0.045-0.055)\GeV^3\qquad\hbox{\rm and}\qquad E_C^{\rm best}=1.4\pm 0.2\GeV.
\end{equation} 
All the results are summarized in the Table 1 , where we compare our results 
with the leading order results obtained in the~\cite{GrYa,Cola,DaHu} and 
some experimental values.  
\begin{table}
\begin{tabular}{|r|c||c|c|c||c|c|}
  \hline
  &Exp.~\cite{UA1}\ and&&&&&\\
  &Pot.model~\cite{Isgur}&\cite{GrYa}&
\cite{Cola}&\cite{DaHu}&L.O.&N.L.O.
\\\hline\hline
  $E_C(\Lambda)$&&$1.20$&$1.2\pm0.1$&$1.20\pm0.15$&$1.2\pm0.2$&$1.2\pm0.2$\\
  $E_C(\Sigma)$&&$1.46$&$1.4\pm0.1$&$1.30\pm0.15$&$1.4\pm0.2$&$1.4\pm0.2$\\
\hline
  $E_R(\Lambda)$&$0.840$($1.040$)&$0.78$&$0.9\pm0.1$&$0.79\pm0.05$
  &$0.8\pm0.2$&$0.92\pm0.2$\\
  $E_R(\Sigma)$&$1.050$($1.250$)&$0.99$&&$0.96\pm0.05$&$1.02\pm0.2$
  &$1.12\pm0.2$\\
  Diff.&$0.190$&$0.21$&&0.17&$0.2$&$0.2$\\
\hline
  $\Delta(\Lambda)$&$0.460$&&&&$0.40\pm0.05$&$0.32\pm0.05$\\
  $\Delta(\Sigma)$&$0.405$&&&&$0.38\pm0.05$&$0.32\pm0.05$\\
\hline\hline
  $F_\Lambda$&&$2.3\pm0.5$&$2.5\pm0.5$&$1.7\pm0.6$&$2.4\pm0.5$&$3.2\pm0.5$\\
  $F_\Sigma$&&$3.5\pm0.6$&$4.0\pm0.5$&$4.1\pm0.6$&$3.4\pm0.5$&$5.2\pm0.5$\\
\hline
\end{tabular}
\caption{All energies like $E_C$, $E_R$ and $\Delta$ are given in $\GeV$,
 the residues in $10^{-2}\GeV^3$. The value of the Borel parameter is 
$T=0.6\GeV$. We use $m_b=4800\MeV$ (resp.\ $4600\MeV$) in the first colomn.}
\end{table}

\section{The Isgur-Wise function}
Next, we consider semileptonic transition $\Lambda_b\to\Lambda_c$. 
The matrix elements of the weak current for 
$\Lambda_b\to\Lambda_c$ at the leading 
order $1/m_Q$ are determined only by Isgur-Wise function 
\begin{equation}
<\Lambda_c|\bar c\Gamma b|\Lambda_b>=\xi(y) \bar u_c\Gamma u_b.
\end{equation}
where $y=v_b\cdot v_c$ and $\Gamma$ is some gamma matrix.
The Isgur-Wise function was calculated in Ref.\cite{GrYa1} 
by using three-point QCD sum rules. 
The slope of Isgur-Wise function at y=1 is 
\begin{equation}\label{slope}
\rho^2=-1.15 \pm 0.2 .
\end{equation} 
The uncertainty is connected mainly with an assumption about continuum model.
We have found that the shape of the Isgur-Wise function nearly coincides 
with the ansatz formula   
\begin{equation}\label{anzatz1}
\xi (y)=\frac2{y+1}exp[(2\rho^2-1)\frac{y-1}{y+1}],
\end{equation}  
with the slope in Eq.(\ref{slope}).
Fig.3 shows the Isgur-Wise function at different energies of the continuum 
($E_c=1.0,1.2,1.4 GeV$).  $1/m_Q$ corections to $\Lambda_b\to\Lambda_c$
transition were discussed recently in \cite{DaHu1}. 
The important $1/m_Q$ effects to 
the decay come only from the weak current expansion. 

\section{Conclusions}

We have calculated one and two-loop anomalous dimensions of baryonic currents. 
We have studied the expansion of the correlator of two heavy 
baryon currents at small Euclidian distances. The radiative corrections to 
the first two Wilson coefficients are calculated. 
The heavy baryons sum rules in 
$\as$ order are derived. 
The predictions for the mass, residues , 
the slope and shape of Isgur-Wise function are presented.

\section{Acknowledgments}

This work was partially supported by the BMBF, FRG, under contract 06MZ566, 
and by the Human Capital and Mobility program under contract CHRX-CT94-0579. 
I would like to thank A.G.~Grozin, J.G.~K\"orner and S.Groote 
for collaboration. I am grateful to organizers of the seminar "Heavy Quark 
Physics".

\begin{figure}
\epsfxsize=15cm
\centerline{\epsffile{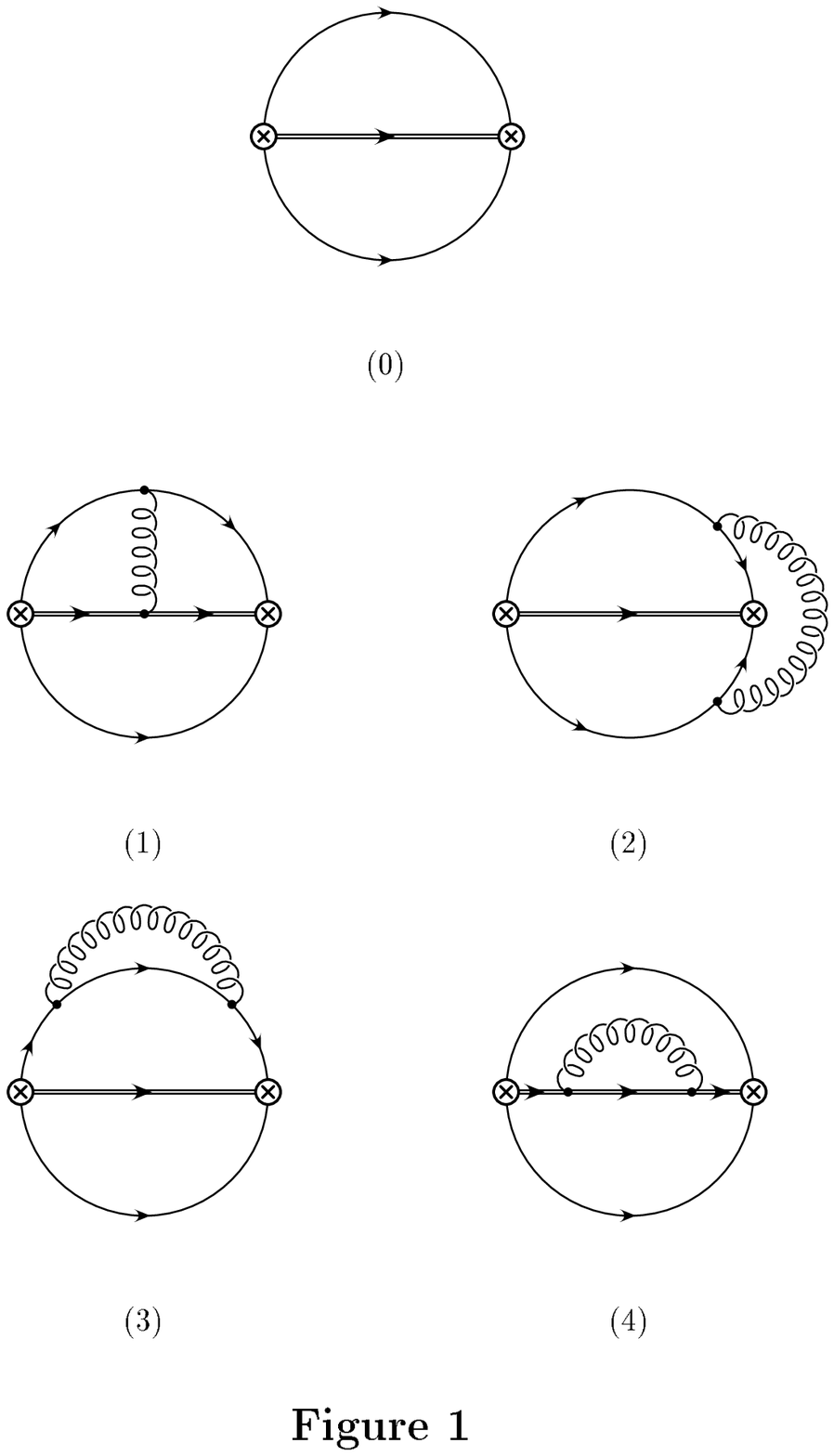}}
\caption[]{Two-loop and three-loop contributions to the correlator 
of two heavy baryon currents. (0) two loop lowest order contribution,
(1)-(4) three loop $O(\alpha_S)$ contributions.}
\end{figure}

\begin{figure}
\epsfxsize=15cm
\centerline{\epsffile{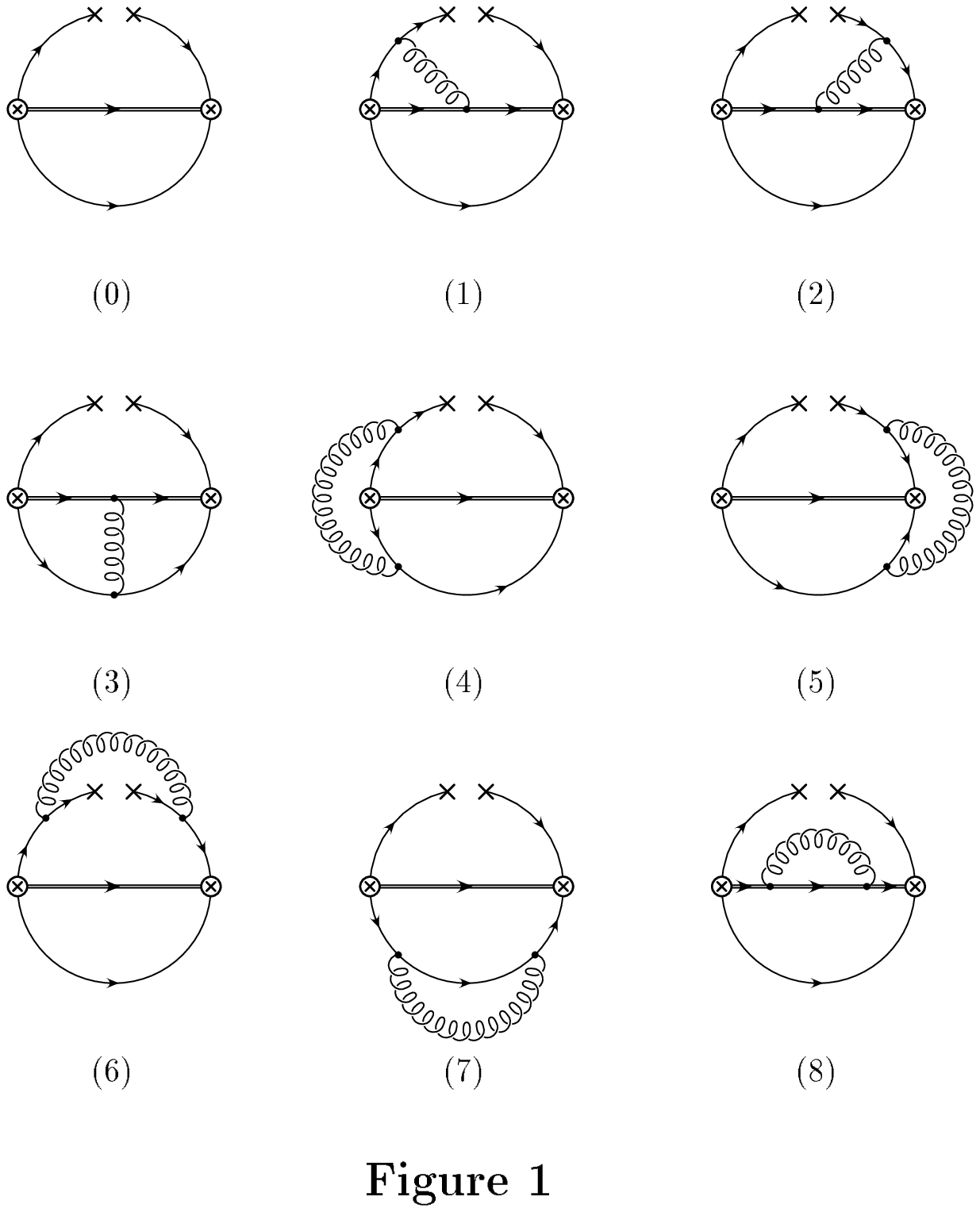}}
\caption[]{One-loop and two-loop contributions to the nondiagonal correlator 
of two heavy baryon currents. (0) one loop lowest order contribution,
(1)-(8) two loop $O(\alpha_S)$ contributions.}
\end{figure}
 
\begin{figure}
\epsfxsize=15cm
\centerline{\epsffile{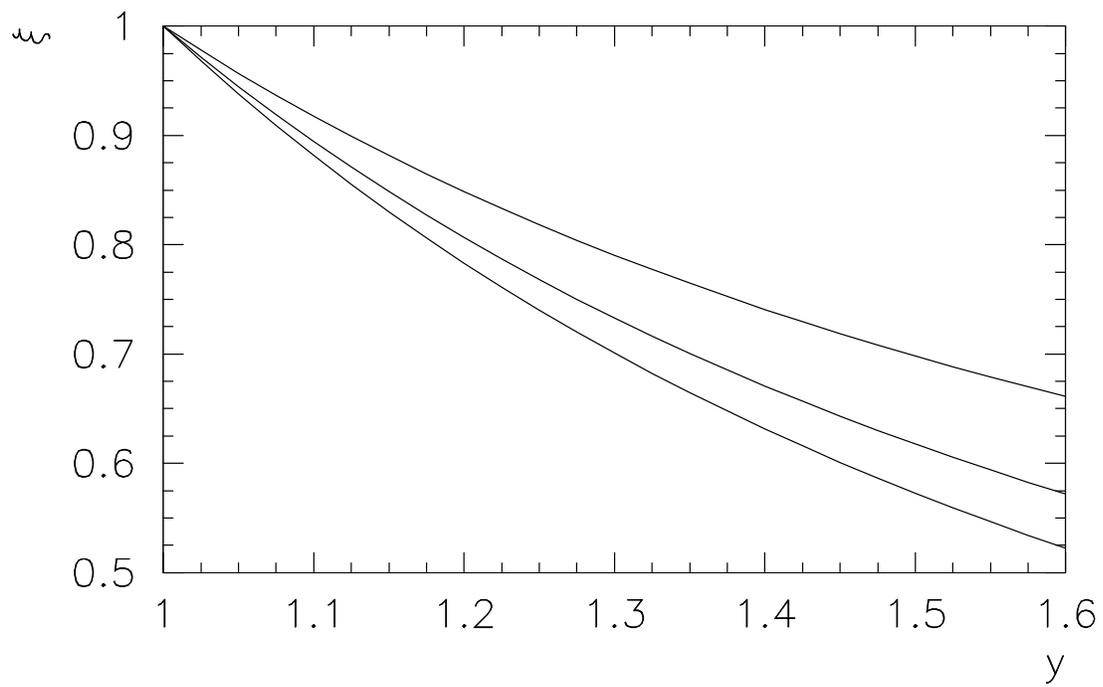}}
\caption[]{The Isgur-Wise function.}
\end{figure}
\end{document}